\documentclass[
reprint,
superscriptaddress,
 amsmath,amssymb,
 pre
]{revtex4-1}

\usepackage{graphicx}
\usepackage[colorlinks=true,allcolors=blue]{hyperref}

\begin{document}

\title{Problems with classification, hypothesis testing, and estimator convergence \\ in the analysis of degree distributions in networks}
\author{Pim van der Hoorn}
\affiliation{Department of Mathematics and Computer Science, Eindhoven University of Technology, Eindhoven, The Netherlands}
\author{Ivan Voitalov}
\affiliation{Department of Physics, Northeastern University, Boston, Massachusetts 02115, USA}
\affiliation{Network Science Institute, Northeastern University, Boston, Massachusetts 02115, USA}
\author{Remco van der Hofstad}
\affiliation{Department of Mathematics and Computer Science, Eindhoven University of Technology, Eindhoven, The Netherlands}
\author{Dmitri Krioukov}
\affiliation{Department of Physics, Northeastern University, Boston, Massachusetts 02115, USA}
\affiliation{Network Science Institute, Northeastern University, Boston, Massachusetts 02115, USA}
\affiliation{Departments of Mathematics and Electrical \& Computer Engineering, Northeastern University, Boston, Massachusetts 02115, USA}

\maketitle

%%%%%%%%%%%%%%%%%%%%%%%%%%%%%%%%%%%%%%%%%%%%%%%%%%%%%%%%%%%%%%%%%%%%%%%%%%%%%%%%%%%%%%%%%%%%%%%%%%%%
%										Brief summary paragraph
%%%%%%%%%%%%%%%%%%%%%%%%%%%%%%%%%%%%%%%%%%%%%%%%%%%%%%%%%%%%%%%%%%%%%%%%%%%%%%%%%%%%%%%%%%%%%%%%%%%%	

In their recent work~\cite{broido2018scale}, Broido and Clauset address the problem of the analysis of degree distributions in networks to classify them as scale-free at different strengths of ``scale-freeness.'' Over the last two decades, a multitude of papers in network science have reported that the degree distributions in many real-world networks follow power laws. Such networks were then referred to as scale-free. However, due to a lack of a precise definition, the term has evolved to mean a range of different things, leading to confusion and contradictory claims regarding scale-freeness of a given network. Recognizing this problem, the authors of~\cite{broido2018scale} try to fix it. They attempt to develop a versatile and statistically principled approach to remove this scale-free ambiguity accumulated in network science literature. Although the paper presents a fair attempt to address this fundamental problem, we must bring attention to some important issues in it:
\begin{enumerate}
  \item {\bf Classification:} Even if a classifier of scale-freeness of a given degree sequence is flawless, there exists an unavoidable and exploitable arbitrariness in applying this classifier to classify networks that are characterized not by one, but by many degree sequences, such as directed, bipartite, multilayer, or temporal networks.
  \item {\bf Correctness:} The classifier of scale-freeness of a given degree sequence used in~\cite{broido2018scale} is not flawless for at least two reasons: 
  \begin{enumerate}
    \item {\bf Hypothesis testing:} The hypothesis testing methodology employed by the classifier makes sense only under very stringent assumptions about the data that are not likely to hold true in any real data.
    \item {\bf Estimator convergence:} Even if the data does satisfy these assumptions, the hypothesis testing classifier yields correct answers only if the power-law exponent estimator employed by the classifier performs sufficiently accurately. However, how large the data size must be for the required accuracy is entirely unknown. 
  \end{enumerate}
\end{enumerate}
Taken together these issues introduce subtle but potentially major defects in the validity of the conclusions in~\cite{broido2018scale}. In the following, we discuss these issues in more detail.

%%%%%%%%%%%%%%%%%%%%%%%%%%%%%%%%%%%%%%%%%%%%%%%%%%%%%%%%%%%%%%%%%%%%%%%%%%%%%%%%%%%%%%%%%%%%%%%%%%%%
%								Point 1: Classification
%%%%%%%%%%%%%%%%%%%%%%%%%%%%%%%%%%%%%%%%%%%%%%%%%%%%%%%%%%%%%%%%%%%%%%%%%%%%%%%%%%%%%%%%%%%%%%%%%%%%

{\bf 1. Classification.} The main result in~\cite{broido2018scale} is the classification of a large collection of real-world complex networks into a complex collection of classes of different strengths of scale-freeness. Here, we do not discuss the definitions of these classes proposed in~\cite{broido2018scale}. What is important is that all of these definitions rely on the outcomes of the classifier from~\cite{clauset2009power}. This classifier classifies a given degree sequence as scale-free or not.

For now, consider this classifier as a black box, and suppose that it is ``ultimately correct.'' Only a small fraction of networks analyzed in~\cite{broido2018scale} are undirected, unipartite, single-layer and static, thus having only one degree sequence. Most networks in~\cite{broido2018scale} are directed, multipartite, multiplex, multilayer, temporal, or combinations of these. Therefore, they have not one but \emph{many\/} degree sequences---as many as the number of snapshots in case of temporal networks, for instance. Applied to each such sequence, the classifier reports whether it is scale-free or not. That is, for most networks, there are not one but \emph{many\/} scale-free/not-scale-free answers from the classifier. Based on these many answers, how can one tell whether a given network as a whole is scale-fee or not?

A significant portion of~\cite{broido2018scale} is an attempt to find an appropriate answer to this question. We believe that this question should not be even pursued because it does not really make much sense. Indeed, it asks to apply a categorization (``scale-free'') of objects of one type (degree sequences) to objects of a different type (networks). Therefore, any attempt to answer this question would introduce unavoidable arbitrariness that can be exploited to tune the overall results to any desirable outcome.

To see why, imagine a universe in which all networks are multilayer with exactly two layers, one of which is classified as scale-free, while the other is not scale-free. If we say that a network is scale-free when at least $50\%$ of its degree sequences are scale-free, then all networks in our universe are scale-free. But if we change our definition just a little bit---a network is scale-free when strictly more than $50\%$ of its degree sequences are scale-free---then we suddenly change our conclusion about the universe to exactly the opposite---our universe has no scale-free networks at all! In both cases we completely ignore and obfuscate the fact that $50\%$ of network layers in our universe are scale-free.

In simple terms, it makes no sense to call a bucket ripe based on any percentage of ripe apples it contains.

%%%%%%%%%%%%%%%%%%%%%%%%%%%%%%%%%%%%%%%%%%%%%%%%%%%%%%%%%%%%%%%%%%%%%%%%%%%%%%%%%%%%%%%%%%%%%%%%%%%%
%									Point 2: Correctness
%%%%%%%%%%%%%%%%%%%%%%%%%%%%%%%%%%%%%%%%%%%%%%%%%%%%%%%%%%%%%%%%%%%%%%%%%%%%%%%%%%%%%%%%%%%%%%%%%%%%

{\bf 2. Correctness.} Above we have assumed that the degree sequence classifier from~\cite{clauset2009power} is ``ultimately correct.'' This is, unfortunately, not the case. Here, we discuss the two most important issues with the classifier. They lie within its two different parts: the hypothesis testing methodology and the power-law parameter estimator.

{\bf 2(a). Hypothesis testing.} The hypothesis that the classifier from~\cite{clauset2009power} tests is whether a given degree sequence (in a given real-world network) is likely to be an i.i.d.\ sample from a pure power-law distribution. A pure power-law distribution is defined as the distribution with the probability mass function $P(k) = C k^{-\gamma}$ for all degrees $k \ge k_{\mathrm{min}}$, where $\gamma > 0$ is the power-law tail exponent, $C$ is the normalizing constant, and $k_{\mathrm{min}}\ge1$ is the minimum degree above which the pure power-law behavior is observed; for $k<k_{\mathrm{min}}$, $P(k)$ can be anything. The two parameters of such a distribution---the power-law exponent $\gamma$ and the minimum degree $k_{\mathrm{min}}$---are estimated within the second part of the classifier by the \texttt{PLFit} estimator, also from~\cite{clauset2009power}.

Similarly to the previous section, consider this estimator as a black box for now, and suppose that it is ``ultimately correct.'' Applied to any degree sequence and relying on the assumption that the degree sequence was indeed sampled i.i.d.'ly from a pure power-law distribution, the estimator always reports its best estimates $\hat{\gamma}, \hat{k}_{\mathrm{min}}$ of the distribution parameters $\gamma, {k}_{\mathrm{min}}$. Given these estimates, the hypothesis testing methodology then proceeds as follows.

First, the hypothesized distribution is set to be the pure power-law distribution with parameters $\gamma=\hat{\gamma}$ and ${k}_{\mathrm{min}}=\hat{k}_{\mathrm{min}}$, and the Kolmogorov-Smirnov (KS) distance~$D^\ast$ between this distribution and the empirical distribution of the given (real-world) degree sequence under investigation is computed for $k\geq{k}_{\mathrm{min}}$. The KS distance between two distributions is the largest distance between their cumulative distribution functions. Then, a large number of artificial degree sequences are sampled i.i.d.'ly from the hypothesized distribution, and for each sequence the KS distance between the sequence and the hypothesized distribution is computed as well, resulting in a sequence of KS distances. Finally, the $p$-value is defined as the fractions of these distances larger than~$D^\ast$. The hypothesis is rejected if $p<0.1$. Indeed, if $p$ is low, then with high probability, an i.i.d.'ly sampled sequence from the hypothesized distribution is KS-closer to this distribution than the given (real-world) sequence of interest. If so, then this sequence is unlikely to be an i.i.d.\ sample from the hypothesized pure power-law distribution.

What makes this hypothesis testing procedure possible in principle is the initial test assumption that the source distribution is a pure power law. This is because pure power laws (known as Zeta distributions in statistics) form a \emph{parametric\/} family of distributions whose parameters $\gamma,k_{\mathrm{min}}$ can be estimated by an estimator. As evident from the exposition above, the described hypothesis testing methodology would fall apart for any \emph{nonparametric\/} (infinite-dimensional) family of distributions, simply because it would require an estimation of an infinite number of parameters from finite data.

However, the same assumption that makes this hypothesis testing methodology possible, also makes it quite unrealistic. This is because there are no known reasons to expect that the complex stochastic processes driving the formation and dynamics of real-world networks would result in pure power-law distributions.
Even if some unknown underlying cause does try to generate pure power laws in a given real-world network, its results are certainly expected to get perturbed by other random interfering processes, noise, measurement inaccuracies, data processing artifacts, and so on.

Pure power laws are not often seen in network \emph{models} either. In fact, we are not aware of any popular network model with pure power-law degree distributions, other than the configuration model where this purity is enforced by hand. Remarkably, generating networks in this model properly and efficiently is quite a challenge~\cite{del-genio2010efficient}. All other popular network models do not produce pure power laws.

The most basic example of that this is indeed the case, is preferential attachment---the ``harmonic oscillator'' of power laws in network science. As was proven two decades ago in~\cite{dorogovtsev2000structure,krapivsky2000connectivity,bollobas2001degree}, when network science was born, the degree distribution in the original plain vanilla version of preferential attachment is not a pure power law but $P(k)=4/k(k+1)(k+2)$, $k=1,2,\ldots$. This distribution belongs to the class of \emph{regularly varying} distributions. This large class of distributions subsumes pure power laws, but also includes all ``impure'' ones. Our review of this 
mathematically well-studied generalization of pure power laws has recently appeared in~\cite{voitalov2019scale}.

If pure power laws are not expected to be present in data, then the described hypothesis testing procedure that looks for them, is expected to yield negative results. These expected results are the main results in~\cite{broido2018scale}, documented succinctly in its title.

That is, the main results in~\cite{broido2018scale} can be summarized as follows. Instead of trying to make rigorous sense out of numerous reports over the last two decades stating that the degree distributions in many real-world networks are \emph{close} to power laws (in the famous but informal $P(k)\sim k^{-\gamma}$ formula), the authors misinterpret these reports as if they report \emph{pure} power laws ($P(k)=Ck^{-\gamma}$). This purity is the key assumption in the described hypothesis testing procedure used to define the proposed scale-free classifications. It is definitely not surprising then that the main result of this misinterpretation of what ``scale-free'' might mean is that ``scale-free networks are rare.'' However, this is not the end of the story.

{\bf 2(b). Estimator convergence.} Thus far we have assumed that the \texttt{PLFit} estimator of $\gamma$ and ${k}_{\mathrm{min}}$ from~\cite{clauset2009power} is ``ultimately correct.'' Indeed, it was very recently shown in~\cite{bhattacharya2020consistency} that \texttt{PLFit} is consistent. Here, \emph{consistency} means that if we draw an increasing number~$n$ of samples from a fixed regularly varying distribution with the exponent~$\gamma$ and then feed them to the \texttt{PLFit}, then the  \texttt{PLFit}'s estimates $\hat{\gamma}$ of $\gamma$ based on these samples tend to $\gamma$ as $n$ tends to infinity. We note that the consistency proof in~\cite{bhattacharya2020consistency} does not directly apply to degree distributions because it applies only to continuous regularly varying distributions. A degree distribution can be made continuous by adding uniform continuous noise to integer degrees, for instance. No \texttt{PLFit} implementation does that, but informal intuition and empirical evidence suggest that the \texttt{PLFit} is likely to be consistent in direct application to discrete regularly varying distributions as well. 

The main problem here is that even though we now have a proof that the \texttt{PLFit} estimates~$\hat{\gamma}$ converge to~$\gamma$ as $n\to\infty$, there are no results whatsoever on the speed of this convergence. That is, it is entirely unknown how large $n$ must be so that the \texttt{PLFit}'s $\hat{\gamma}$ lies within a given accuracy window around the true~$\gamma$.

Since we do not know how large the data must be so that the \texttt{PLFit} is sufficiently accurate, it is quite possible that a given (real-world) network is so small that the \texttt{PLFit}'s estimate~$\hat{\gamma}$ is very far from the true~$\gamma$. If such an inaccurate estimate is then fed to the hypothesis testing methodology described above, it results in a large KS distance $D^\ast$ simply because we got the $\gamma$ wrong, which inevitably leads to a low $p$-value. As a result, the hypothesis is wrongfully rejected. 

We note that this issue was well understood in~\cite{clauset2009power}, whose authors wrote, ``\ldots one prefers large statistical samples when attempting to verify hypotheses such as these,'' or ``\ldots $p$-values should be treated with caution when $n$ is small.'' However, the authors of~\cite{broido2018scale} do not mention this issue at all.

The \texttt{PLFit} is not an exception in terms of unknown convergence speeds. There are plenty of consistent estimators of power-law exponents developed in statistics. Almost nothing is known about the speed of convergence of any of those estimators. Therefore, there is no way to place any rigorous confidence intervals on their estimates, unless the data satisfies some strong assumptions that can never be verified in reality~\cite{voitalov2019scale}. This lack of control over estimation accuracy and error makes the \texttt{PLFit}, as well as all other power-law exponent estimators, impossible to use in any hypothesis testing procedure, including the one described above. What can we do then in such daunting conditions?

\begin{figure}[!]
\centering
\includegraphics[scale=0.21]{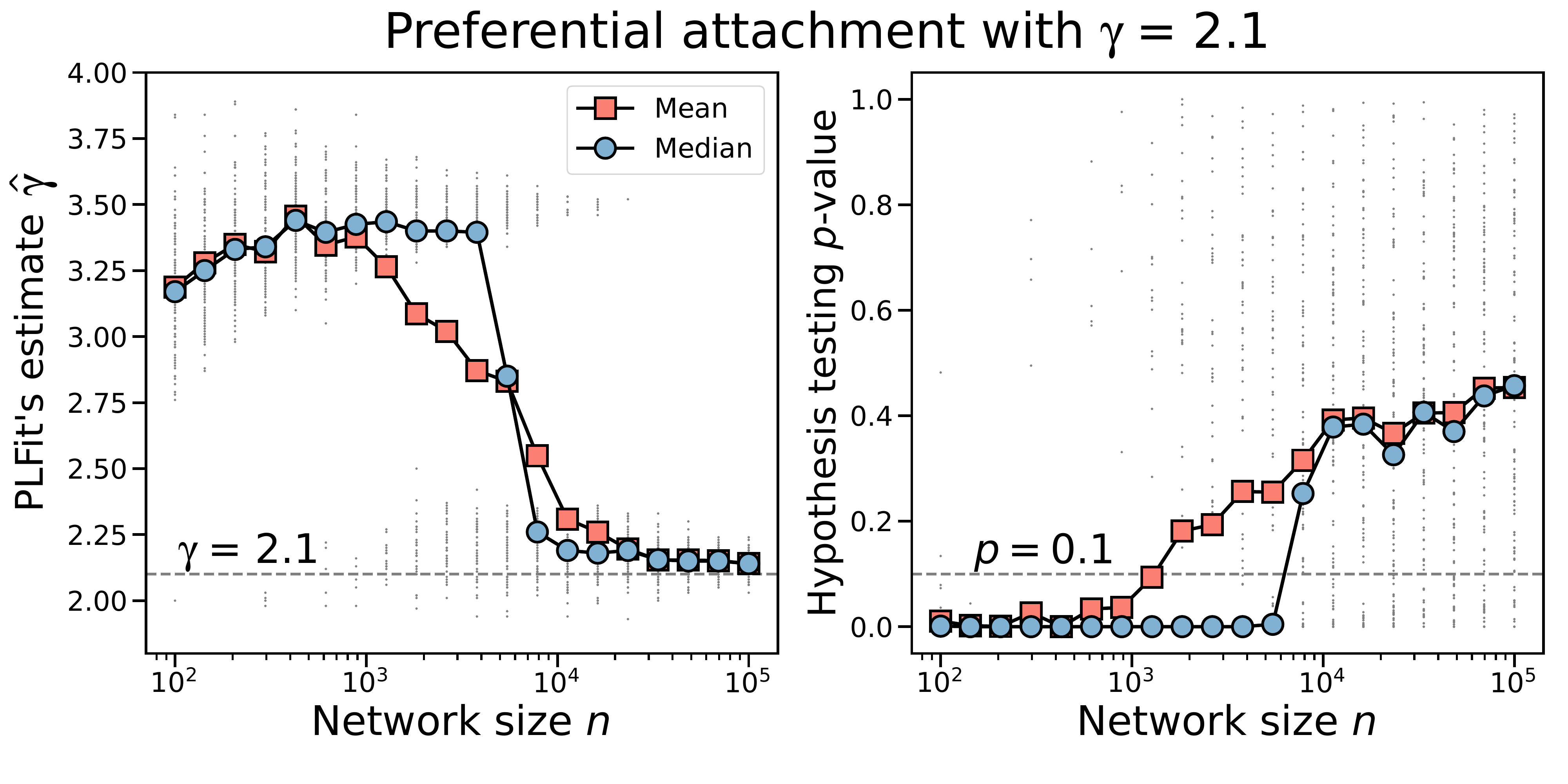}
\caption{{\bf The convergence speed of the \texttt{PLFit} estimator~\cite{clauset2009power} applied to preferential attachment networks.} For each network size~$n$, 100 random networks of this size are grown according to the preferential attachment algorithm based on redirection~\cite{krapivsky2001organization} with the power-law exponent set to $\gamma = 2.1$. For each network, the \texttt{PLFit} estimator is applied to its degree sequence, and the resulting estimates are then fed to the hypothesis testing procedure from~\cite{clauset2009power} that computes the $p$-value. The hypothesis that the degree sequence is sampled from a power law is rejected if $p<0.1$. The left and right panels display the estimates of~$\gamma$ and the $p$-values, respectively. The dots show these data for each network, while the squares and circles show the means and medians for each network size~$n$.
}
\label{fig:p_value_experiment}
\end{figure}

Since there is no theory behind how fast \texttt{PLFit}'s estimates converge to the true values, the best we can do to get a feeling about the convergence speed is to apply the estimator to a sequence of degree sequences of increasing size known to converge to a power law~\cite{voitalov2019scale}. In Figure~\ref{fig:p_value_experiment} we do so for our power-law ``harmonic oscillator''---preferential attachment.

We see in Figure~\ref{fig:p_value_experiment} that the \texttt{PLFit} estimates~$\hat{\gamma}$ do appear to converge to the true value $\gamma = 2.1$ as the network size grows, suggesting that the \texttt{PLFit} converges not only on continuous distributions, but also on discrete ones. However, we also see that this convergence is not monotone, and the point where the  \texttt{PLFit}'s estimation accuracy improves significantly coincides with the point where the median of the $p$-value crosses the $0.1$ threshold, which happens for network sizes about~$10^4$. In other words, in the considered preferential attachment networks with more than $10^4$ nodes, the methodology from~\cite{clauset2009power} works as one would expect. However, for smaller-size networks this procedure breaks down, returning wrong estimates of~$\gamma$ and hence small $p$-values, leading to erroneous rejection of the hypothesis that we are dealing with power laws.

We have just seen that for the methodology from~\cite{clauset2009power} to work correctly, networks must be larger than~$10^4$ in a very controlled lab environment with relatively ``clean'' (albeit not pure) power laws in preferential attachment. What can we then say about real-world networks where power laws, if any, are expected to be much ``dirtier'' in a completely uncontrolled and unknown way? How large should such networks be for a given estimation accuracy?

Unfortunately, we cannot say anything at all in this regard for the following two related reasons. First, we usually have no idea what the true distribution producing the observed degree sequence in a given real-world network is. Second, the speed of \texttt{PLFit}'s convergence likely depends very strongly not only on key parameters of this distribution, such as~$\gamma$, but also on its very fine details, especially the details of its deviation from a pure power law. One thing is crystal clear though: networks must be ``very large.'' However, since there is no theory behind the convergence speed, there is no way to tell how large this ``very large'' is.

How large are the real-world networks considered in~\cite{broido2018scale}? The authors report this information only indirectly in Figure~1 in~\cite{broido2018scale}. Careful examination of this figure shows that the fractions of networks with sizes below $10^3$ and $10^4$ are about $50\%$ and $70\%$, respectively. That is, a majority of real-world networks considered in~\cite{broido2018scale} are actually smaller than~$10^4$, the size needed for synthetic preferential attachment networks to be correctly diagnosed as scale-free using the methodology employed in~\cite{broido2018scale}. For the reasons above, it is not surprising then that this methodology finds the considered real-world networks to be ``rarely scale-free.'' For the same reasons, there are no guarantees that these findings reflect reality.

%%%%%%%%%%%%%%%%%%%%%%%%%%%%%%%%%%%%%%%%%%%%%%%%%%%%%%%%%%%%%%%%%%%%%%%%%%%%%%%%%%%%%%%%%%%%%%%%%%%%
				%							Summary
%%%%%%%%%%%%%%%%%%%%%%%%%%%%%%%%%%%%%%%%%%%%%%%%%%%%%%%%%%%%%%%%%%%%%%%%%%%%%%%%%%%%%%%%%%%%%%%%%%%%

{\bf Summarizing,} 
the conclusions in~\cite{broido2018scale} rely on the following major assumptions, organized in the order of logical dependency:
\begin{enumerate}
  \item the network is large enough for the \texttt{PLFit} estimator to yield sufficiently accurate estimates that do not break the hypothesis testing procedure,
  \item the (unknown) network formation process produces pure power-law degree sequences so that the hypothesis testing procedure can be employed, and the resulting $p$-values can be trusted, and
  \item the proposed network classification scheme, which relies on as many of such $p$-values as the number of the degree sequences that the network has, makes perfect sense.
\end{enumerate}
Our main point is that it is impossible to asses the validity of the first innermost assumption in application to any real data, so that whether the second and third assumptions are valid does not really matter. However, we have also argued that they are likely to be invalid as well. Taken together, these points imply that extreme care must be taken when it comes to interpreting the conclusions in~\cite{broido2018scale}.

\bibliographystyle{prx-bibstyle}
\bibliography{references}

\end{document}